\documentstyle[12pt,epsfig]{article} % Simple LaTeX, use with DINA4 as follows
\newlength{\dinwidth}                       
\newlength{\dinmargin}                      
\def\lsim{\mathrel{\rlap{\lower4pt\hbox{\hskip1pt$\sim$}}
    \raise1pt\hbox{$<$}}}                % less than or approx. symbol
\def\gsim{\mathrel{\rlap{\lower4pt\hbox{\hskip1pt$\sim$}}
    \raise1pt\hbox{$>$}}}                % greater than or approx. symbol

\def\g{\gamma}
\def\cts{\cos\theta^{\ast}}

\def\xgo{x_\g^{OBS}}

\def\xpo{x_p^{OBS}}

\def\ETAJ{\eta^{jet}}
\def\ETJ{E_T^{jet}}

\def\MJJ{M_{JJ}}

\begin{document}

\begin{flushright}
DESY 96-191 \\
UCL/HEP 96-04
\end{flushright}

%\vspace*{1cm}
\begin{center}  \begin{Large} \begin{bf}
High $p_T$ Charm Photoproduction \\
  \end{bf}  \end{Large}
  \vspace*{5mm}
  \begin{large}
G. Abbiendi$^a$ J. M. Butterworth$^b$, R. Graciani$^c$\\ 
  \end{large}
\end{center}
$^a$ Deutsches~Elektronen-Synchrotron~DESY, 
     ~Hamburg,~Germany\\
$^b$ Dept. of Physics and Astronomy, University College London,
     London, UK.\\
$^c$ Universidad Autonoma de Madrid, Spain.\\
\begin{quotation}
\noindent
{\bf Abstract:}
The expected rates for charm-tagged jet photoproduction are evaluated
for a number of tagging procedures, and some of the physics potential
is discussed. Charm in jets is tagged using $D^*$'s,
$\mu$'s, or tracks from secondary vertices which might be identified
in a microvertex detector.  We find high expected event rates,
leading to the possibility of placing strong constraints on the
kinematics of charm production and on the gluon content of the proton
and the charm content of the photon.
\end{quotation}
\section{Introduction}

At HERA energies interactions between almost real photons (of
virtuality $P^2 \approx 0$) and protons can produce jets of high
transverse energy ($\ETJ$). A significant fraction of these jets are
expected to arise from charmed quarks. The presence of a `hard' energy
scale means that perturbative QCD calculations of event properties can
be confronted with experiment, and hence the data have the potential
to test QCD and to constrain the structures of the colliding
particles.

\begin{figure}
\psfig{figure=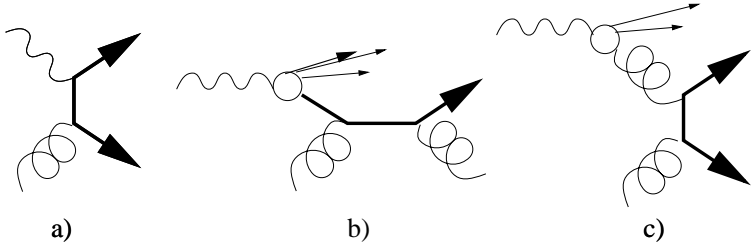,height=10cm,angle=270} 
\caption{\label{f:diagrams} \it Examples of charm photoproduction at HERA: a)
Direct; photon-gluon fusion, b) Resolved; single excitation of charm in 
the photon, c) Resolved; gluon-gluon fusion. The
charm and anticharm quarks are indicated by the bold lines.}
\end{figure}

At leading order (LO) two processes are responsible for jet
production. The photon may interact directly with a parton in the
proton or it may first resolve into a hadronic state. In the first
case all of the photon's energy participates in the interaction with a
parton in the proton. In the second case the photon acts as a source
of partons which then scatter off partons in the proton. Examples of
charm production in these processes are shown in Fig.~\ref{f:diagrams}.

The possibility of experimentally separating samples of direct and
resolved photon events was demonstrated in~\cite{Z2}, and in~\cite{Z4}
a definition of resolved and direct photoproduction was introduced
which is both calculable to all orders and measurable. This
definition is based upon the variable
\begin{equation}
\xgo = \frac{ \sum_{jets}\ETJ e^{-\eta^{jet}}}{2yE_e},
\label{xgoeq}
\end{equation}
\noindent where the sum runs over the two jets of highest $\ETJ$.
$\xgo$ is thus the fraction of the photon's energy participating in
the production of the two highest $\ETJ$ jets. This variable is used
to define cross sections in both data and theoretical calculations.
High $\xgo$ events are identified as direct, and low $\xgo$ events as 
resolved photoproduction.

Charm-tagged jet cross sections have several advantages over untagged
jet cross sections. Knowledge of the nature of the outgoing parton
reduces the number of contributing subprocesses and thus simplifies
calculations and the possible extraction of parton densities,
including the charm content of the photon and the gluon content of the
proton. Detailed studies of the dynamics of charm production should
provide a stringent test of the QCD calculations. In addition, in the
case that charm decays are fully reconstructed, the outgoing momenta
provide an alternative to calculating the event kinematics from jet
momenta, which could provide a useful {\it model independent}
examination of the uncertainties coming from non-perturbative
fragmentation effects.

Here we briefly examine the event rates and distributions obtainable
with high luminosities using the three charm tagging methods described
below. We use the HERWIG~5.8~\cite{HERWIG} Monte Carlo 
simulation, including multiparton interactions~\cite{MI}, along
with simple cuts and smearing to mimic the expected gross features of
detector effects. We define our starting sample by running the $k_T$
jet algorithm~\cite{kt} on the final state particles of HERWIG (after
the decay of charmed particles) and demanding at least two jets with
transverse energy $\ETJ > 6$GeV and absolute pseudorapidity $|\ETAJ| <
2$. In addition we demand $P^2 < 4$GeV$^2$ and 135~GeV $< W_{\gamma p}
< 270$~GeV. This is a kinematic region in which dijet cross sections
have already been measured at HERA~\cite{Z4}.

According to HERWIG, the total cross section for heavy flavour ($b$ or $c$) 
jets in this kinematic region is 1900~pb$^{-1}$ (1000~pb$^{-1}$) for direct
(resolved) photoproduction, using the GRV94 LO proton parton distribution
set~\cite{GRV94} and the GRV LO photon parton distribution set~\cite{GRVph}.  

There is some evidence~\cite{Z4,RG} that these LO calculations may
underestimate the cross section, particularly the resolved cross
section, by a factor of around two. On the other hand, the dominant
LO subprocess for resolved charm production is predicted to be
excitation of charm from the photon. This expectation is not
fully reliable: the charm content in the photon is presently
overestimated in the available parton distribution sets as they assume
only massless quarks. If quark masses are included one would expect
the resolved charm cross section to be considerably {\it smaller} than
the number we are quoting here. Its measurement will be an important
topic in its own right.

\section{Charm tagging methods}

\subsection{$D^*$ tagging method}

Currently, the reconstruction of $D^*$ is the only method used to tag
open charm by the HERA experiments in published data~\cite{dstar}. 
$D^*$ are tagged by
reconstructing the $D^0$ produced in the decay 
$D^{*\pm} \rightarrow D^0 + \pi^\pm$ and the mass difference $\Delta(M)$ between the
$D^*$ and the $D^0$.

The overall tagging efficiency for the $D^*$ method is given in
table~1, along with the expected number of events after an integrated
luminosity of 250~pb$^{-1}$.  For this study we have demanded  
a $D^*$ with $p_T > 1.5$~GeV and $|\eta| < 2$, and assumed that for 
these $D^*$ the efficiency of
reconstruction is 50\%.  The decay
channels used are $D^* \rightarrow D^0 + \pi \rightarrow ( K + \pi ) +
\pi$ and $D^* \rightarrow D^0 + \pi \rightarrow ( K + \pi \pi \pi ) +
\pi$.  
A signal/background ratio of around 2 is estimated, although this (as well
as the $D^*$ reconstruction efficiency) will depend upon
the understanding of the detectors and cuts eventually achieved
in the real analysis, which cannot be simulated here.

\subsection{$\mu$ tagging method}

The capability of the $\mu$ tagging method has been evaluated using a
complete simulation of the ZEUS detector~\cite{ZEUS} based on the
GEANT package \cite{GEANT}. The method itself develops previous work
\cite{GA} in which a measurement of the total charm photoproduction
cross section was obtained in the range $60 < W < 275$~GeV. Muons are
tagged requiring a match between a track in the ZEUS central tracking
detector pointing to the interaction region and a reconstructed
segment in the inner muon detectors (which lie about one metre away, outside
the uranium calorimeter).   

The position and the direction of the reconstructed segment are used
to determine the displacements and deflection angles of its
projections on two orthogonal planes with respect to the extrapolated
track.  These quantities are distributed according primarily to the
multiple Coulomb scattering within the calorimeter. In comparison the
measurement errors are negligible and have not been taken into
account.  With this approximation and a simple model accounting for
the ionization energy loss of the muon through the calorimeter, a
$\chi^2$ has been defined from the four variables.
%The position of the hits and the direction
%of the reconstructed segment measured on the muon detectors are used
%to determine the displacements and deflection angles on two orthogonal
%planes with respect to the extrapolated track. These quantities are
%distributed according primarily to the multiple Coulomb scattering
%within the calorimeter, for true muons. In comparison the measurement
%errors are negligible for our purposes. With this approximation and a
%simple model accounting for the ionization energy loss of the muon
%through the calorimeter a $\chi^2$ has been defined from the four
%variables.  
%Displacement and deflection on one plane are obviously
%correlated.  
The cut on the $\chi^2$ has been chosen to keep 90\% of
the events with a reconstructed true muon in large Monte Carlo charm
samples and checked in selected data samples. The results are
contained in table~1.

\subsection{Tagging using secondary vertices}

If a high resolution microvertex detector is installed close to the interaction
region, the tagging of charm by looking for secondary vertices inside jets
becomes practical. For this study we have simulated three example methods
(`A', very tight cuts and `B', looser cuts and `C', very loose cuts) 
as follows:

\begin{itemize}

\item Look at all stable charged tracks 
which have transverse momentum 
$p_T({\rm track}) > 500$~MeV and $|\eta({\rm track})| < 2$ and 
which lie within $\delta R = \sqrt{(\delta\phi)^2+(\delta\eta)^2} < 1.0$
of the centre of either of the two jets, and 

\item Assume a (Gaussian) impact parameter 
resolution for these tracks of $180~\mu$m in $XY$ and $Z$ independent
of momentum and angle. This corresponds to the design value
of the H1 vertex detector~\cite{H1} for tracks with momentum
500~MeV at $90^o$.

\item Demand at least two tracks which have impact parameters
displaced by 3$\sigma$ (condition A)
or one track with an impact parameter displaced by 3$\sigma$
(condition B) or 2$\sigma$ (condition C) from the primary vertex.

\end{itemize}

The results are given in table~1. We note that an enriched  sample
of $b$ quarks could be obtained by using very tight tagging conditions
in a microvertex detector.

\begin{table}[t]
\centering
\begin{tabular}{|c|c|c|c|c|c|c|}  \hline
Tagging  & \multicolumn{3}{c|}{ Direct }& \multicolumn{3}{c|}{ Resolved } \\ \cline{2-7} 
Method   & Efficiency & N(events) & Sig./Bkgd & Efficiency & N(events) & Sig./Bkgd \\ \hline
$D^*$    & 1.4\% &  6500 ( 9\% b) &  $\approx 2$& 0.7\% & 1700 ( 4\% b)& $\approx 1$ \\
$\mu$    & 7.3\% & 34000 (20\% b) &  2.0      & 3.4\%  &  8400 (10\% b)& 0.3       \\
Vertex A & 2.3\% & 11000 (63\% b) &  76       & 1.0\%  &  2500 (34\% b)& 8         \\ 
Vertex B & 10\%  & 47000 (33\% b) &  3.4      & 6.0\%  & 15000 (17\% b)& 0.5       \\ 
Vertex C & 37\%  &170000 (17\% b) &  0.9      & 32\%   & 79000 (6\% b) & 0.2       \\ \hline
\end{tabular}
\caption{\it Estimated tagging efficiencies, 
signal to background ratio and total numbers of expected 
signal events for various tagging methods after an integrated
luminosity of 250~pb$^{-1}$.  The efficiencies given are the ratios of
good events which are tagged to all good events.  `Good events' are
$ep \rightarrow 2$ or more jets with $\ETJ \ge 6$~GeV, $|\ETAJ| < 2$,
for virtualities of the exchanged photon less than 4 GeV$^2$ in the
range 135~GeV $< W_{\gamma p} < 270$~GeV and where
one or more of the outgoing partons from the hard subprocess was a
charm or beauty quark. The fraction of the signal events which
are from $b$ quarks is also given.}
\end{table}

\section{Physics Potential and Conclusions}

High luminosity running at HERA will provide large samples of jets
containing heavy quarks. These jets can be identified using muons or
$D^*$ with efficiencies of a few percent and signal-to-background
ratios of around 2. In addition there is the possibility of
identifying the electron channel for semi-leptonic decays, which we
have not considered here but which could be very effective at these
high transverse energies. The presence of a high resolution vertex
detector would enormously enrich the tagging possibilities, allowing
improved signal-to-noise ratios and/or improved efficiencies (up to
around 35\%) depending upon the details of the cuts and
reconstruction. Combining the tagging methods we have studied here can
be expected to give still more flexibility in the experimental
selection and cross section measurement.

With the samples of several tens of thousands of charm-tagged jets
thus obtainable, jet cross sections can be measured over a wide
kinematic range.  For the signal events selected by the vertex method
B, various distributions are shown in Fig.2.  From the $\xgo$
distribution (Fig.2a) we see that the resolved photon component,
whilst suppressed relative to the direct component compared to the
untagged case~\cite{Z4}, is significant. This component is largely due
to the charm content in the GRV photon parton distribution set.
Measurement of this cross section can be expected to constrain the
charm content of the resolved photon and the implementation of the
$\gamma \rightarrow c\bar{c}$ splitting in the perturbative evolution.
The boson gluon fusion diagram dominates for the high-$\xgo$ range and
this cross section is sensitive to the gluon content of the proton in
the range $0.003 < \xpo < 0.1$, where $\xpo = \frac{ \sum_{jets}\ETJ
e^{\eta^{jet}}}{2E_p}$ is the fraction of the proton's energy manifest
in the two highest $\ETJ$ jets (Fig.2b).  The $\MJJ$ distribution is
shown in Fig.2c, where $\MJJ = \sqrt{ 2 E_T^{jet1}
E_T^{jet2}[\cosh(\eta^{jet1} - \eta^{jet2}) - \cos(\phi^{jet1}
-\phi^{jet2})]}$ is the dijet invariant mass.  For $\MJJ > 23$~GeV the
dijet angular distribution~\cite{Z5} $|\cts| =
|\tanh(\frac{\eta^{jet1} - \eta^{jet2}}{2})|$ is unbiased by the
$\ETJ$ cut. As shown in Fig.2d the angular distributions of high and
low $\xgo$ should differ strongly, due to the underlying bosonic
(gluon) or fermionic (quark) exchange processes~\cite{Z5}. The
measurement of such a distribution should confirm that the dominant
charm production process in direct photoproduction was photon-gluon
fusion. In addition it will determine whether excitation of charm from
the incoming particles or gluon-gluon fusion is the dominant
production mechanism in resolved photoproduction.

\begin{figure}
\hspace{2cm}
\psfig{figure=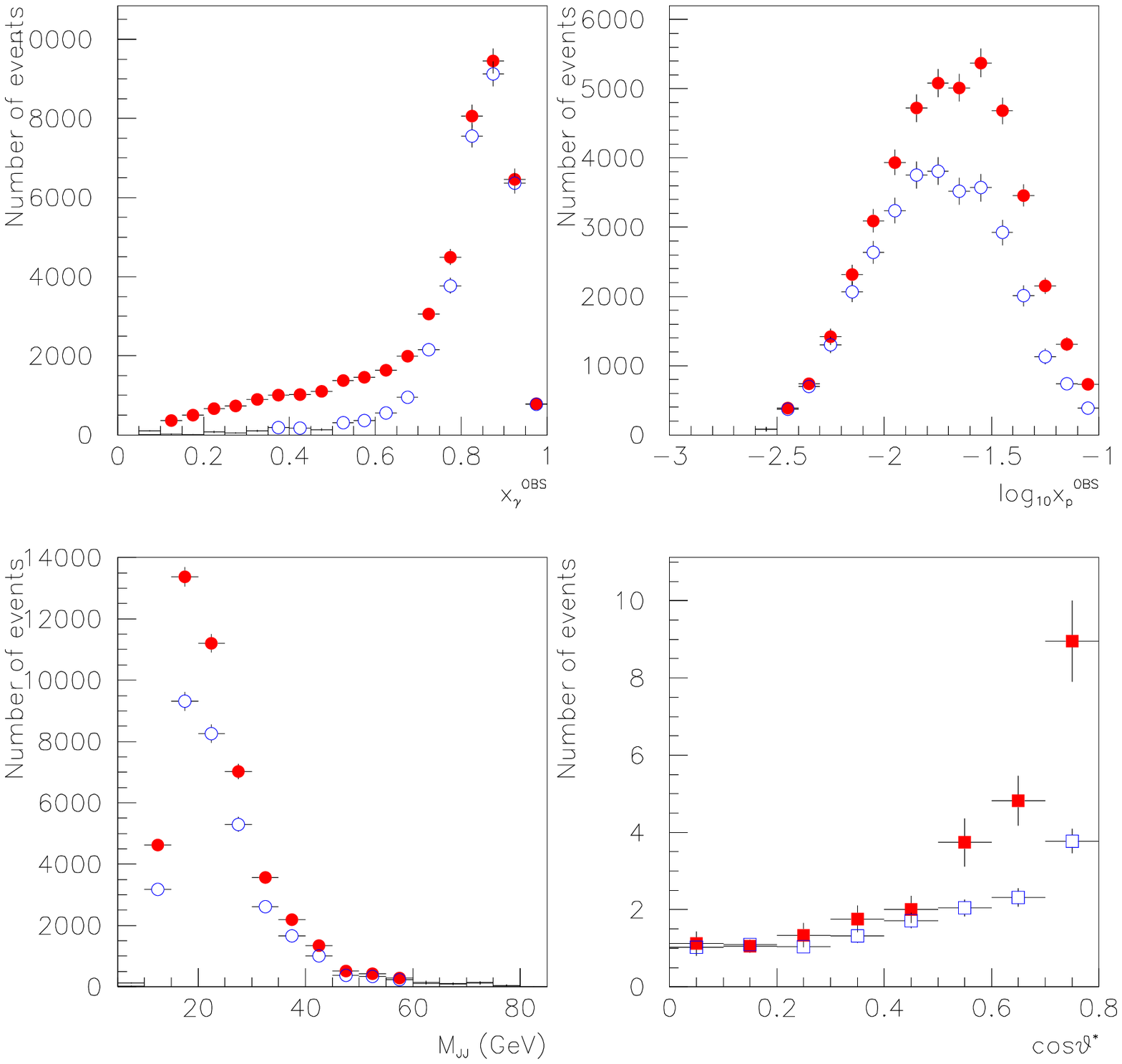,height=14cm} 
\caption{\label{f:dist} \it a) $\xgo$, b) $\xpo$, c) $\MJJ$, and d) $\cts$.
In a), b) and c), clear circles are the LO direct only, solid dots are the
full sample. The normalisation is to 250~pb$^{-1}$. 
In d) the solid squares are the $\xgo < 0.75$ sample and
the clear squares are the $\xgo > 0.75$ sample. Both samples are normalised to
one at $|\cts| = 0$ and the error bars have been scaled to
correspond to the statistical uncertainty expected after 250~pb$^{-1}$.}
\end{figure}

\section*{Acknowledgements}
It is a pleasure to thank U. Karshon, D. Pitzl, A. Prinias,
S. Limentani and the members and conveners of the working group for
discussions and encouragement.

\end{document}